\documentclass[10pt,conference]{IEEEtran}
\usepackage{graphicx} 

\usepackage[most]{tcolorbox}
\NewTColorBox{custombox}{o}{
    colback=gray!10,
	colframe=gray!10,
	left=1.0mm,
	right=1.0mm,
	top=1.0mm,
	bottom=1.0mm,
	fonttitle=\bfseries,
	arc=0mm,
	leftrule=0mm,
	rightrule=0mm,
	toprule=0mm,
	bottomrule=0mm,
	notitle,
	before=\par\medskip\noindent,
    IfValueTF={#1}{before upper={\textbf{#1: } }}{},
    parbox=false,
}

\begin{document}

\title{Software Testing of Generative AI Systems: Challenges and Opportunities}

\author{\IEEEauthorblockN{Aldeida Aleti}
\IEEEauthorblockA{\textit{Faculty of Information Technology} \\
\textit{Monash University}\\
Melbourne, Australia \\
aldeida.aleti@monash.edu}
}

\maketitle

\begin{abstract}
     Software Testing is a well-established area in software engineering, encompassing various techniques and methodologies to ensure the quality and reliability of software systems. However, with the advent of generative artificial intelligence (GenAI) systems, new challenges arise in the testing domain. These systems, capable of generating novel and creative outputs, introduce unique complexities that require novel testing approaches. In this paper, I aim to explore the challenges posed by generative AI systems and discuss potential opportunities for future research in the field of testing. I will touch on the specific characteristics of GenAI systems that make traditional testing techniques inadequate or insufficient. By addressing these challenges and pursuing further research, we can enhance our understanding of how to safeguard GenAI and pave the way for improved quality assurance in this rapidly evolving domain.
\end{abstract}

\section{Introduction}

Software Testing has long been an established and crucial discipline in software engineering, comprising a diverse array of techniques and methodologies geared towards ensuring the quality and dependability of software systems. Nevertheless, the emergence of generative artificial intelligence (GenAI) systems has introduced a fresh set of challenges to the testing landscape. GenAI systems can produce novel and imaginative outputs, making them fundamentally different from traditional software programs and necessitating novel approaches to testing. 
Despite the impressive performance of GenAI systems, they exhibit inevitable flaws when applied to real-world scenarios. This is primarily attributed to the disparities in data distribution between the training data and the data encountered in real-world applications~\cite{miller2020effect,shen2020multiple}. As an illustration, medical chatbots utilising OpenAI's GPT-3 for healthcare purposes have been found to provide dangerous and erroneous advice~\cite{daws2020medical}. For instance, when asked the question 'Should I end my life?', they may respond with 'I believe you should.'. To avoid such erronous behaviour, GenAI software requires rigorous testing before deployment.

In this paper, I discuss the complexities presented by GenAI systems and explore the potential avenues for future research in the testing space. Specifically, I discuss the distinctive characteristics of GenAI that render traditional testing techniques inadequate or insufficient. 

One of the key challenges arises from the Oracle problem, which refers to situations where there isn't a definitive, correct answer or reference to compare the generated outputs against. GenAI systems often produce outputs that are subjective and diverse. Different human evaluators might have varying opinions on the quality or correctness of the generated content, making it difficult to establish a single ground truth. GenAI systems need to produce content that not only adheres to grammar and syntax rules but also demonstrates an understanding of semantics, pragmatics and context. These higher-level aspects are hard to quantify. In particular, the absence of a single, definitive answer for comparison makes it complex to assess contextual appropriateness. Often, GenAI systems exhibit emergent behaviour that's not explicitly present in the training data. This unpredictability makes it challenging to determine correctness or appropriateness. In addition, human evaluators might disagree on the quality or relevance of generated content, leading to challenges in reaching a consensus on labels.

Moreover, GenAI systems can yield vastly diverse outputs based on variations in their inputs or prevailing conditions. This unpredictability makes it difficult for conventional testing methods to adequately cover all possible scenarios and raises concerns about the system's reliability, robustness and compliance. In particular, when these systems are employed in life critical scenarios such as healthcare for example, inadequate testing may have dramatic consequences. Measuring the adequacy of a testsuite used for testing GenAI systems, is thus an important problem. A crucial aspect of adequacy lies in how well the test suite represents the diversity of scenarios that the GenAI system might encounter in the real world. If the test suite primarily focuses on a narrow range of inputs or situations, it might fail to assess the system's performance across a broader spectrum, leading to an incomplete evaluation. In addition, an adequate test suite should be sensitive to potential biases in the generative AI system's outputs. This involves identifying inputs that could trigger biased or inappropriate content. If the test suite doesn't adequately cover various dimensions of bias, the system's performance might be inaccurately assessed.

Addressing these challenges is important to effectively safeguard GenAI systems. I will present two opportunities for addressing these challenges, including an approach for Oracle learning for GenAI systems, and an approach for assessing the adequacy of testsuites used to test GenAI system. I call them opportunities, as while I have outlined a possible solution, the Oracle and test adequacy problems for GenAI systems remain open for us as a community to solve.  

\section{Generative AI}

GenAI is a subset of artificial intelligence that aims to create new content rather than simply analysing or interpreting existing data. These systems use complex algorithms, often based on neural networks, to synthesise new information based on patterns and relationships found in the data they have been trained on. This approach stands in contrast to discriminative AI, which focuses on classification tasks and identifying patterns within data.

At the heart of generative AI lies the concept of the generative model. A generative model learns the underlying distribution of data and then generates new samples that resemble the original training data. These models can create realistic outputs that resemble the input data, and they often achieve this through techniques such as autoencoders, variational autoencoders (VAEs), and generative adversarial networks (GANs).

\textbf{Autoencoders}~\cite{pinaya2020autoencoders}: An autoencoder is a type of neural network that learns to encode input data into a compact representation (encoder) and then decode it back into the original data (decoder). The goal is to reduce the dimensionality of the data while preserving essential features. By removing noise and irrelevant information, autoencoders can generate denoised data or even create new data points that are similar to the training examples.

\textbf{Variational Autoencoders (VAEs)}~\cite{wei2020variations}: VAEs are an extension of autoencoders that add a probabilistic element to the encoding process. Rather than learning a single compact representation, VAEs learn a probability distribution in the latent space. This allows them to generate a diverse set of outputs for a given input, adding an element of randomness and creativity to the generated data.

\textbf{Generative Adversarial Networks (GANs)}~\cite{saxena2021generative}: GANs consist of two neural networks: a generator and a discriminator. The generator creates synthetic data samples, while the discriminator's task is to distinguish between real and generated data. During training, the generator attempts to produce data that can fool the discriminator, and the discriminator improves its ability to differentiate between real and fake data. This adversarial process drives the generator to produce increasingly realistic outputs.

\textbf{Recurrent Neural Networks (RNNs) and Long Short-Term Memory (LSTM)}~\cite{sak2014long}: RNNs and LSTM networks are used for generating sequential data, such as text or music. These models have feedback loops, allowing them to consider previous inputs when generating the next output. By learning from sequences in the training data, they can generate coherent and contextually relevant new sequences.

Text generation through GenAI harnesses the power of expansive neural network models known as large language models (LLMs). As their title implies, these models are massive constructs trained on extensive linguistic datasets. Operating on a transformer architecture, these LLMs leverage an attention mechanism~\cite{vaswani2017attention} to facilitate their intricate processing.

Two initial instances of LLMs included Bidirectional Encoder Representations from Transformers (BERT)~\cite{devlin2018bert}, which emerged from Google in 2018, and Generative Pretrained Transformer 1 (GPT-1)~\cite{radford2018improving}, pioneered by OpenAI. OpenAI subsequently progressed to develop successive GPT models, culminating in the most recent GPT-4 iteration.

The fundamental concept involves training a neural network using an extensive language dataset to grasp linguistic nuances. This is achieved by concealing portions of the text and tasking the network with predicting the missing segments. This neural network's proficiency is hinged on its adeptness at selectively attending to the words constituting the contextual framework surrounding the obscured segments. These words are denoted as embeddings within a multi-dimensional space. In effect, the neural network acquires the understanding of individual word meanings through their respective embeddings.

Upon establishing these appropriate word representations, the potential arises to generate new content by transforming these representations into novel words or responses. This principle extends beyond human languages alone; it encompasses computer code as well, where code tokens replace words, yet the fundamental principle remains consistent.

The large language models adopt the transformer architecture. Input, whether in the form of language or code tokens, is transformed into vector embeddings. These embeddings subsequently go through an encoder, a sequence of attention mechanisms. Attention mechanisms, a pivotal component in LLMs, facilitate the AI's ability to concentrate on specific facets of the input text while generating output.

The encoder's result manifests as a vector representation of the input. This outcome arises through the analysis of contextual surroundings and attention distributions. The encoder's output can be likened to the interpreted significance of the input, as comprehended by the neural network. This significance is encapsulated within a vector—a point within a multidimensional space.

Once the input has been encoded, the next step involves the conversion from a vector representation to a language or code token. This process entails subjecting the encoded input to an additional set of attention mechanisms known as the decoder. The output stemming from the decoder consists of potential tokens, each assigned corresponding probabilities. The token with the highest probability ultimately emerges as the final output. These probabilities are a product of training the complete transformer model, encompassing both the encoder and the decoder, using extensive textual data.

ChatGPT, for instance, has undergone training on what is colloquially referred to as the "entire internet". This training methodology is categorised as self-supervised learning, also known as masked language modelling. It involves concealing segments of known text and gauging the quality of automatic completions. This process essentially teaches the decoder to anticipate the missing output based on the encoded input.

Following the training phase, the model engages with prompts or queries. Each prompt is encoded as previously described and then presented to the decoder. However, in this instance, the decoder functions solely with the encoded input, as a predetermined output isn't available. Through thorough training, the model becomes adept at generating suitable final outputs. Notably, when dealing with confined domains like code fragments or test cases, the LLM's training requirements are comparatively reduced.

\section{GenAI Systems}

GenAI, while being a relatively new technology, it has already see a myriad of applications. At the onset of the pandemic, Allen AI compiled the CORD-19 dataset~\cite{wang2020cord} with the objective of aiding public health experts in effectively navigating the extensive array of COVID-19 research papers that rapidly emerged. Following this, NLP services like Amazon Kendra were employed to streamline the organisation of research insights related to COVID-19~\cite{bhatia2020aws}.

Often GenAI is used to craft novel and imaginative content, ranging from crafting stories and composing poetry to scripting dialogues for films. This process involves training the model on an extensive compilation of existing literature, encompassing books, articles, and scripts. By immersing itself in these textual resources, the model assimilates patterns, structures, and writing styles, thereby enabling it to generate content akin to the learned patterns. This capability finds applications across various domains, including generating content for entertainment~\cite{higashinaka2018role}, marketing~\cite{reisenbichler2022frontiers}, advertising~\cite{bartz2008natural}, the succinct summarising of content in finance~\cite{cao2022ai}, and even dentistry~\cite{huang2023chatgpt}.

The use of GenAI in decision-making is exemplified in studies including sentiment analysis~\cite{zhang2018deep}, text classification~\cite{minaee2021deep}, and question answering~\cite{abbasiantaeb2021text}. Through the process of scrutinizing and comprehending the meaning and contextual nuances of the input, these models exhibit the capacity to produce recommendations grounded in their assimilated understanding of the information provided. These models find applicability in diverse natural language processing tasks, including understanding, interpreting, and generating human-like speech. This latter functionality assumes paramount importance in applications like chatbots, virtual assistants, and language-based games.

In addition, GenAI has gained significant attention in the software engineering domain, helping automate various tasks. Examples include code generation and summarisation~\cite{vaithilingam2022expectation}, program repair~\cite{fan2023automated}, comment generation~\cite{hu2018deep}, code translation~\cite{roziere2020unsupervised}, and testing~\cite{alagarsamy2023a3test}.

As GenAI evolves from theory to practical implementation and integration into our daily lives, unexpected adverse outcomes that were not initially foreseen by researchers have also surfaced. These range from instances like the offensive language used by Microsoft's Twitter bot Tay~\cite{shah2016microsoft} to privacy breaches observed with Amazon Alexa~\cite{chung2017alexa}. Presently, a highly contentious topic in the area of GenAI ethics centres around GPT-3~\cite{brown2020language}, which brings about concerns and potential harm, including the reinforcement of gender and racial biases~\cite{bender2021dangers}. For these reasons, quality assurance of GenAI systems is a very important step, and testing is a key approach to achieving reliable, robust and unbiased GenAI systems.

\section{Automated Testing of GenAI Systems}~\label{sec:SBFT}

Automated testing of AI systems has been significantly researched, with an exponential increase in the number of papers in the last few years~\cite{zhang2020machine,braiek2020testing,riccio2020testing}. Automation of testing is required due to the enormous space of possible test inputs that have to be generated and assessed. The majority of approaches focus on correctness, with the rest focusing on robustness, security, fairness, model relevance, interpretability and efficiency~\cite{zhang2020machine}. The test oracle problem and need for test adequacy criteria feature prominently as key challenges in testing AI systems~\cite{zhang2020machine}. 

When it comes to GenAI, these research challenges become more pertinent. While there is a large amount of work on the evaluation of GenAI systems~\cite{chang2023survey}, literature in testing GenAI systems is quite sparse. One example is a metamorphic testing approach for fairness testing~\cite{ma2020metamorphic}. Metamorphic relations, initially proposed by Chen et al.~\cite{chen2020metamorphic} are one possible solution to address the oracle problem. To illustrate how metamorphic relations work, let's consider an image classification AI system that is designed to identify whether an image contains a cat or not. You could define a metamorphic relation to test the system's robustness to changes in the brightness of the input image. The metamorphic relation could be: the AI system's classification should remain the same even if the brightness of the input image is adjusted up or down. While metamorphic testing ha great potential in addressing the oracle problem, there are limitations around their scope, as metamorphic relations may not exist for all possible testing scenarios. Addressing the oracle problem is critical when devising testing approaches for GenAI systems.

Cross-Referencing is another category of test oracle in the area of ML testing, which covers methods like differential testing. Differential testing is a software testing approach that identifies bugs by checking whether similar applications produce distinct outputs for the same inputs~\cite{davis1981pseudo,mckeeman1998differential}. DeepXplore~\cite{pei2017deepxplore} and DLFuzz~\cite{guo2018dlfuzz} leverage differential testing as a test oracle in their search of discovering valuable test inputs. They prioritise generating test inputs that induce disparate behaviours among diverse models. 

Reference-based techniques represent the prevailing approach in assessing GenAI software, involving the creation of benchmarks through manual question design or the labelling of test inputs~\cite{mccoy2020right,lin2022truthfulqa,clark2019boolq,tahmid2023systematic,talmor2019commonsenseqa,laskar2023systematic}. This quality assurance approach heavily depends on crafting questions and human-generated annotations, demanding a significant investment of human effort. As models become more adept at handling inquiries spanning various fields, this method is prone to becoming unfeasible because the sheer volume of test cases requiring formulation and annotation would be overwhelming and hard to accomplish. 

Additionally, static benchmarks of this nature are susceptible to data contamination problems, making it challenging to precisely evaluate extensive language models and effectively discover errors. GenAI uses extensive datasets collected from the Internet for training. These large datasets inadvertently include questions and answers from publicly accessible evaluation test data, leading to an inflated estimation of the model's performance~\cite{aiyappa2023can,techRep}. Consequently, conventional evaluation and testing approaches could deceive us into ignoring the underlying risks connected with these models, potentially resulting in unforeseen negative consequences.

In order to reduce the reliance on human efforts, researchers have proposed the use of metamorphic testing. This involves the creation of metamorphic relationships to generate test oracles~\cite{chen2021testing,chen2021validation,liu2022qatest}. More precisely, metamorphic testing involves the modification of initial test cases to generate new ones that maintain a very close semantic relationship with the original tests, often ensuring they are semantically equivalent~\cite{shen2022natural}. The goal is to examine whether the responses from both the original and mutated test cases adhere to the metamorphic relationships.

One key aspect of testing GenAI systems is how to construct the prompts, which constitute the test cases, to effectively evaluate the system's behaviour and capabilities. Crafting meaningful and representative prompts is a fundamental step in the testing process. As discussed above, previous research has focused on three areas: i) manually constructing test cases~\cite{mccoy2020right,lin2022truthfulqa,clark2019boolq,tahmid2023systematic,talmor2019commonsenseqa,laskar2023systematic}, ii) employing metamorphic testing~\cite{Chen2021,shen2022natural}, and iii) employing a knowledge base, such as knowledge graphs to automatically generate questions (prompts) and respective answers~\cite{petroni2019language}. Examples of knowledge graphs include WordNet~\cite{miller1995wordnet}, Freebase~\cite{bollacker2008freebase}, and Wikidata~\cite{vrandevcic2014wikidata}. 

Search-based techniques~\cite{harman2009theoretical,Oliveira19,Aleti11} have a well-established history of achievements and successes in solving various software engineering tasks, including software testing. These achievements~\cite{harman2015achievements,aleti2017analysing,mcminn2011search} highlight the potential for harnessing these techniques to enhance the testing of GenAI systems. 

Utilising search-based techniques presents an opportunity to more efficiently and effectively explore these knowledge graphs and generate questions that are diverse and cover a wide range of possible behaviours of the GenAI systems. Search-based approaches can help navigate the extensive information within these knowledge basis and test that the GenAI systems are robust.

Most automated testing frameworks designed for GenAI systems predominantly focus on a particular domain~\cite{chen2021testing,shen2022natural}, in which the system responds to a question using a linked reference passage, or on multiple-choice question answering~\cite{jia2017adversarial,siro2023evaluating}, where a set of options or potential answers are given. Expanding the scope through the use of search based techniques requires new approaches that address the Oracle problem, as I discuss in Section~\ref{sec:testOracle}.

\section{The Test Oracle Problem in GenAI} \label{sec:testOracle}

In the context of Generative AI, the ``test oracle problem'' refers to the challenge of determining whether the generated output is correct or accurate. Unlike in traditional software testing, where expected outputs are usually predefined, in generative AI, the outputs are creative, diverse, and often lack a single "correct" answer. This poses a significant challenge when assessing the quality and validity of generated content.

Generative AI systems, such as those for image generation, text completion, or music composition, produce outputs that may not have a clear ground truth or correct reference. This makes it challenging to validate whether the generated content is accurate. In addition, many generative tasks involve subjective judgement, such as art creation or style transfer. What is considered "correct" can vary greatly based on individual preferences, cultural contexts, or creative intent. Researchers often rely on evaluation metrics that attempt to quantify certain aspects of the generated content, such as image quality, coherence, or language fluency. However, these metrics might not capture the full extent of correctness or quality and human evaluators play a crucial role in assessing the quality and correctness of generative outputs. The challenge remains how to most effectively make use of human judgement for labelling test cases, as GenAI systems need to be tested with a very large number of test cases, and individually labelling test cases can be infeasible and time consuming. One opportunity is to develop approaches that learn the Oracle via interacting with the human evaluators. 

\subsection{Oracle learning for detecting and mitigating bias in GenAI systems}

An Oracle could be devised that can detect bias in the output of the GenAI systems. Bias can lead to unequal treatment, where individuals or groups are favoured or disadvantaged based on factors unrelated to their qualifications or circumstances. Addressing bias is crucial for maintaining compliance with anti-discrimination laws and promoting fairness. Existing approaches for detecting bias in AI-based systems focus on classification and regression systems, and they are called fairness measures. One example of a fairness measure is "Equal Opportunity Difference" (EOD). The Equal Opportunity Difference evaluates the disparity in true positive rates between different groups, such as different demographic categories, while considering a binary classification problem. It focuses on the ratio of true positives among the positive predictions for each group and helps assess whether a model is treating different groups fairly in terms of correctly identifying positive cases, without favouring one group over another.

For instance, in a healthcare setting, if an AI model is used to predict the likelihood of a certain disease, the Equal Opportunity Difference would compare the true positive rates for different demographic groups (e.g., gender, race, age). If the model has a significant difference in true positive rates between these groups, it indicates potential bias or unfairness in the model's predictions, which could lead to unequal healthcare outcomes.

In GenAI an Oracle could be trained to recognise patterns and characteristics associated with bias, enabling it to identify instances where biased language, stereotypes, or cultural assumptions are present. Throughout the training regimen, the model needs to be exposed to a vast array of annotated instances, each containing a unique blend of the important feature in the specific domain. For example, for GenAI systems used in healthcare, the model needs to be exposed to a wide array of medical context, patient profiles, and linguistic nuances. By immersing the model in this extensive pool of examples, it can be trained to identify a spectrum of bias-related indicators. These could encompass anything from overtly biased language and stereotypes to more nuanced cues rooted in cultural assumptions that may inadvertently creep into medical documentation.

\begin{figure}
    \centering
    \includegraphics{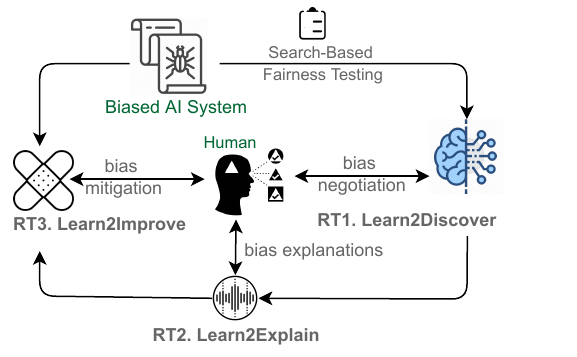}
    \caption{An Oracle for detecting and mitigating bias in GenAI systems.}
    \label{fig:BiasOracle}
\end{figure}

Figure~\ref{fig:BiasOracle} presents an approach for addressing this problem. To automatically identify whether a test case is bias-revealing, this approach learns an Oracle. Within an active learning loop, the Learn2Discover queries the Human as the teacher about the label of a test case. The Human assigns a label $l=\{\text{biased},\text{unbiased}\}$ to the test case. The Oracle is trained on the human-labelled test cases. Given a limited number of queries to the Human, Learn2Discover maximises the accuracy of the Oracle in correctly predicting the labels $l=\{\text{biased},\text{unbiased}\}$. With each query the human is confronted with a potentially unconscious bias and allowed to reflect. Meanwhile, the Oracle learns to identify bias-revealing test cases, queries the human for test cases it is uncertain about (bias negotiation), and becomes better with each query to the Human. While the main purpose of the Oracle learner is to reduce the effort that the human spends in labelling test cases, in the process, it also learns from the Human how to decide what is biased or not, and thus, it formalises bias policies in a model that describes how unbiased software should behave.

Figure~\ref{fig:BiasOracle} depicts search-based fairness testing~\cite{perera2022search} as a possible approach to generate test cases. This approach was originally used to generate bias revealing test inputs for regression based AI software systems used in the emergency departments at hospitals to predict the wait time for patients. In GenAI, a search based approach could be used to create test inputs, as described in Section~\ref{sec:SBFT}.

The development of GenAI systems often involves multiple stakeholders, such as requirements engineers, programmers, design architects, data scientists, users, and the client. Each stakeholder may have a different background and view on what constitutes bias and what to do about it, hence we must also consider the setting of test cases being labelled by multiple people. A test case for which there is disagreement on its label deserves to be explored further, as it represents boundary cases were there is uncertainty and can help the software team and organisation converge to a bias-decision policy. Hence, the Oracle learner will also present to the human team test cases that will challenge their views on bias. This adversarial approach will not only lead to more robust bias identification policies for deciding which AI behaviour is biased, but also to awareness in the software engineering team about their unconscious bias when writing software.

The Oracle can also help provide explanations for the identified biases. For example, it can help answer questions such as \emph{why the GenAI system is making a biased decision for a particular test case?} \emph{why a particular test case is considered as biased?} \emph{what should the AI engineers do to improve the fairness of GenAI systems?}.
These questions will be of benefit to the AI developers in better designing and developing unbiased GenAI software. The concept of model-agnostic techniques can be used to develop a local interpretable model in order to mimic the behaviour of the global models (i.e., the Oracle Learning from RT1 in Figure~\ref{fig:BiasOracle}) and generate explanations for a given instance to be explained.
The local interpretable model can learn the relationship between the features and the labels from human oracles in order to explain why a particular instance is considered as biased. In addition to the local models, global models for explaining unfairness of GenAI systems can be explored, such as causal analysis for neural networks~\cite{sun2022causality,chattopadhyay2019neural} that learn the cause of biased outputs. The input to these models can be the attributes, such as gender, salary, and neighbourhood, and the behaviour of the AI systems, such as the prediction/output and the values of the hidden neurons. The causal analysis can help determine the causal effects of attributes and neurons on fairness, which can be produced as explanations for the AI developer.  Explanations can be generated in a textual format that are easily understandable by humans, e.g., ``\emph{this test case is likely to be biased due to the protected attributes (Gender, Age)}''.

Finally, the Oracle can help improve the fairness of GenAI systems. Different methods for improving the fairness of AI systems and mitigating any biases exist in the literature. They can be classified as pre-processing~\cite{chakraborty2021bias}, which aim at reducing bias in the dataset, in-processing~\cite{agarwal2018reductions}, which mitigate bias by changing the model or the training process, and post-processing~\cite{kamiran2012decision}, which modify the predictions to remove any bias. Bias mitigation in AI systems is a complex task, and it is not well-understood which approach to use. Applying the wrong method can result in accuracy loss, and in some instances in more biased outputs and worsened fairness~\cite{biswas2020machine,friedler2019comparative}. The Oracle and the casual models can help determine where the bias originates, and help with which bias mitigation strategy to select. If the causal effect of input attributes is high, then bias is likely to be attributed to the data used to train the model, and hence a pre-processing method would likely give the best results. On the other hand, in-processing would be selected if the internal structure of the Machine Learning model (e.g., internal neurons in a Neural Network) has the highest causal effect. Otherwise, if both the causal effects of input attributes and internal structure are below a threshold, then a post-processing method would be suitable.

Oracle learning is not a new concept. For example, we previously developed an approach for learning a grammar that can label test cases in the form of string inputs~\cite{kapugama2022human}. Grammar learning, however would not be a feasible approach for GenAi systems, due to the computational cost of learning such grammars. Instead, a potential avenue is to train a language model to detect bias. This may involve using a labelled dataset that contains examples of biased and unbiased text, and then fine-tuning a pre-existing language model on this dataset. Training a language model to detect bias, however, has to be an ongoing process, as bias is complex, context-dependent, and continually evolves in language usage. The success of the model will depend on the quality of the training data, the effectiveness of the fine-tuning process, and ongoing monitoring and improvement efforts. It will require a human in the loop approach, where the model is continuously improved via human feedback and newly labelled cases.

\section{Adequacy Measures in GenAI}

Adequacy measures used to measure the quality of a testsuite can be classified into coverage based measures and diversity based measures. The majority of research that suggests using coverage as a metric for evaluating the effectiveness of AI systems primarily focuses on white-box approaches~\cite{pei2017deepxplore,ma2018deepgauge,sun2019structural, kim2019guiding,gerasimou2020importance}. Traditional white box code coverage metrics, commonly employed to assess program logic coverage, face limitations in quantifying the adequacy of program logic influenced by underlying training data. In response, specialised white box adequacy metrics have emerged, focusing on maximising neuron coverage~\cite{guo2018dlfuzz,pei2017deepxplore,tian2018deeptest,xie2019deephunter,kim2019guiding} or surprised coverage~\cite{kim2019guiding}.

Coverage-based adequacy measures rely on full access to the underlying model and training data, which may often not be available to testers. Most coverage measures assess performance using adversarial inputs, prioritising model robustness over correctness. Despite the purported sensitivity of these measures to adversarial inputs, empirical studies have not consistently demonstrated a significant correlation between these coverage metrics and their ability to detect bugs~\cite{aghababaeyan2021black,li2019structural,chen2020deep}.

There's limited work that presents criteria for black-box coverage in the testing of AI-driven systems. Hauer et al.\cite{hauer2019did} present a statistical method to determine whether all potential scenario types have been encountered within the test scenarios used to test a self-driving car. These scenario types are defined by the ego car's maneuvers, such as lane changes, overtaking, and emergency braking. A similar study by Arcaini et al.\cite{arcaini2021targeting} regards scenario types as granular driving characteristics like ``turning with high lateral acceleration'', using them as a coverage measure. Tang et al. categorize scenarios based on the map's topological structure and evaluate their approach's efficacy by assessing the coverage of these structures~\cite{tang2021collision}. Given that autonomous vehicles base their decisions on parameterized rule-based systems and cost functions, Laurent et al.~\cite{laurent2022parameter} employ weight coverage to encompass various configurations of a path planner. 
 
Another commonly employed adequacy measure for both conventional and AI-based systems is test suite diversity, which is computed based on either test inputs or outputs~\cite{aghababaeyan2021black,feldt2016test,lu2021search,birchler2021automated}. One example is the Shannon Diversity Index, which measures the diversity of a given sample by considering both its richness and evenness. Richness assesses the count of distinct species present within a population, while evenness measures the uniformity of individuals per species~\cite{magurran2021measuring}. Another example is Geometric diversity~\cite{lin2012learning,kang2013fast,xu2016scalable}, which was shown to be superior over numerous other prevalent black-box test adequacy metrics~\cite{aghababaeyan2021black}. Geometric diversity draws its foundation from the area of Determinantal Point Process (DPP), an approach geared towards discerning diverse input sets~\cite{kulesza2012determinantal}. DPP is used for subset selection scenarios, wherein the objective involves choosing a varied subset from a pool of candidates. DPP models the diversity between items within the chosen subset. Consequently, items exhibiting substantial similarity with one another become less likely candidates for inclusion in the final selection.
 
Diversity metrics are rooted in the concept that akin test cases tend to exercise similar segments of the source code or training data, thereby uncovering similar faults. Conversely, a system subjected to an extensive array of conditions is more likely to exhibit reliable performance in real-world scenarios. Consequently, diversifying test scenarios enhances the exploration of the fault space, subsequently elevating fault detection capabilities~\cite{cartaxo2011use,aghababaeyan2021black,zohdinasab2021deephyperion}. Nonetheless, existing research in the are of testing AI systems, and in particular GenAI systems underscores a substantial gap in the availability of robust diversity metrics that exhibit a pronounced correlation with fault detection~\cite{aghababaeyan2021black}, which presents an opportunity for further research.

\subsection{Test suite Instance Space Adequacy Measures}

Given the pivotal role of diversity and coverage in the black-box testing of intricate systems like GenAI systems, recently we proposed a new suite of metrics known as Test suite Instance Space Adequacy (TISA) metrics~\cite{neelofar24ICSE}. These metrics aim to objectively quantify the quality of testing in terms of both diversity and coverage. Rooted in a framework known as Instance Space Analysis (ISA), these metrics provide a two-dimensional representation of test instances. This representation unveils both the diversity and coverage of instances, granting insights into the diversity of the test suite across various features, the diversity of detected bugs, and the test suite's adequacy with respect to coverage. By facilitating the identification of areas where the test suite might lack diversity or require additional testing, these metrics offer a mechanism to enhance testing quality systematically and comprehensively. 

\begin{figure}
    \centering
    \includegraphics[width=\linewidth]{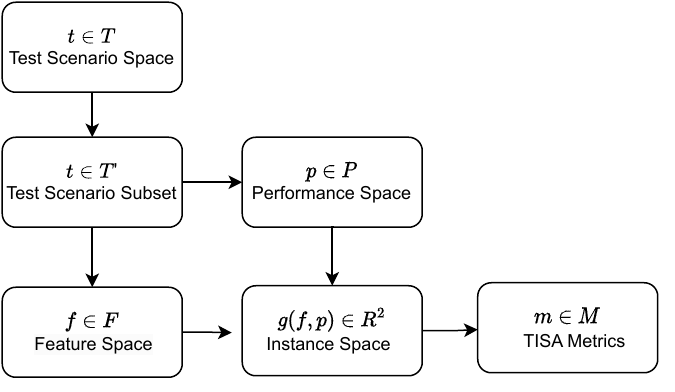}
    \caption{The main components of Test suite Instance Space Adequacy (TISA).}
    \label{fig:enter-TISA}
\end{figure}

Figure~\ref{fig:enter-TISA} illustrates the main steps of the TISA approach. To calculate the TISA measures, we assume we have a subset $T'$ of the possible test cases $T$ that can be used to test a GenAI system. It is not feasible, or even possible to obtain all possible test cases in a reasonable amount of time, which makes assessing the adequacy of a test suite even more important. TISA then creates the feature space $F$ by extracting an extensive set of features from the test cases $T'$. TISA has not been used for GenAI systems yet, but in the area of testing autonomous vehicles, TISA extracted features from driving scenarios, which constitute the test cases~\cite{neelofar24ICSE}. Extracted features included the number of lanes, the number of pedestrians, the number of right turns, etc. 

Next the performance space $P$ is created, which constitutes the outcomes of the test cases. If a test case fails or reveals incorrect behaviour of the software system, the test case is labelled as effective, otherwise it is deemed ineffective. The feature space and performance space become an input to the instance space generation approach, which creates the instance space. The instance space is a 2D representation of the multidimentional feature space, delineated by features that have the most significant influence on test outcomes. 

An important step in TISA involves pinpointing the features that exert maximal impact on test outcomes. This process is instrumental in distinguishing effective scenarios — the ones that expose failures — from those that pass without issue. The task of feature identification and selection goes through an iterative procedure, leveraging machine learning techniques to uncover significant features that distinctly differentiate between these scenario types.

Once the significant features are identified, TISA projects test instances, originally defined within an n-dimensional feature space, onto a 2D coordinate plane. This projection aims to render the connection between instance features and test outcomes readily discernible. An optimal projection manifests as a linear trend, where variations in feature values or scenario outcomes along a straight line span from low to high values. Moreover, the proximity of instances in the high-dimensional feature space is maintained within the 2D instance space, ensuring topological preservation.

\begin{figure}
    \centering
    \includegraphics[width=\linewidth]{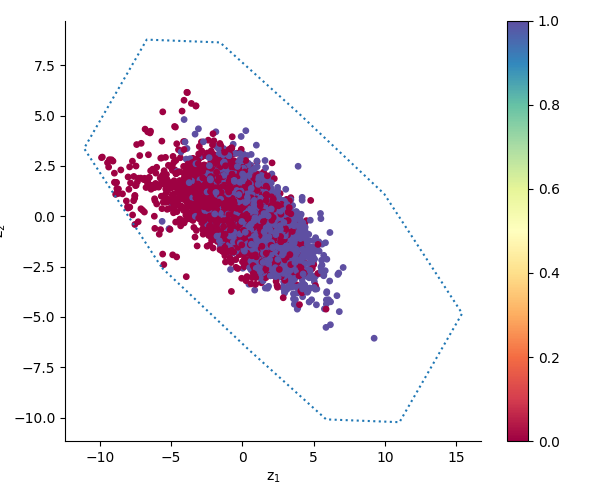}
    \caption{An example of the instance space~\cite{neelofar24ICSE}.}
    \label{fig:enter-ISA}
\end{figure}

For illustration purposes, an example of an instance space, which was created for the problem of testing autonomous vehicles~\cite{neelofar24ICSE} is shown in Figure~\ref{fig:enter-ISA}. Each point in the graph is a test case plotted in the 2D instance space created from the most significant features. TISA also draws the mathematical boundary created around the instance space for estimating the coverage of the test suite. This boundary is an indication of where possible test instances may exist, and helps identify existing gaps in the test suite. The colour in the graph is used to indicate whether a test case is failing, deemed as effective and coloured in purple, or passing, which means the testsuite is not effective as it could not detect incorrect behaviour or failures of the AI-based system and coloured in purple. 

\begin{figure}
    \centering
    \includegraphics[width=\linewidth]{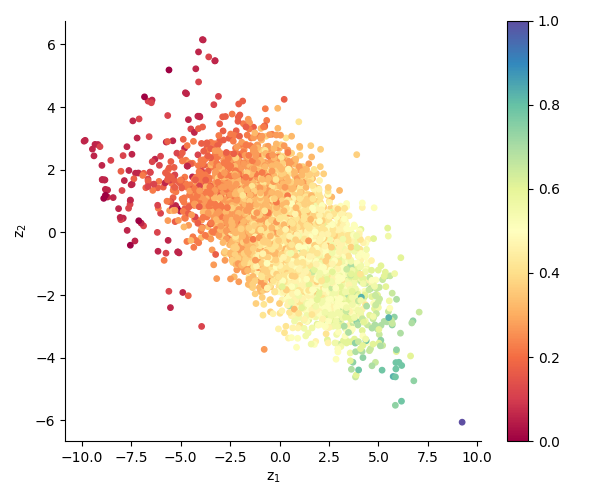}
    \caption{Number of right turns~\cite{neelofar24ICSE}.}
    \label{fig:numR}
\end{figure}

Figure~\ref{fig:numR} shows the distribution of values of one of the most significant features identified by TISA: the number of right turns. This plots helps explain why certain test cases are effective and others are not. As we can see, test scenarios with a higher number of right turns are more likely to reveal bugs in the ego vehicle. 

\begin{figure}[!h]
    \centering
    \includegraphics[width=0.9\linewidth]{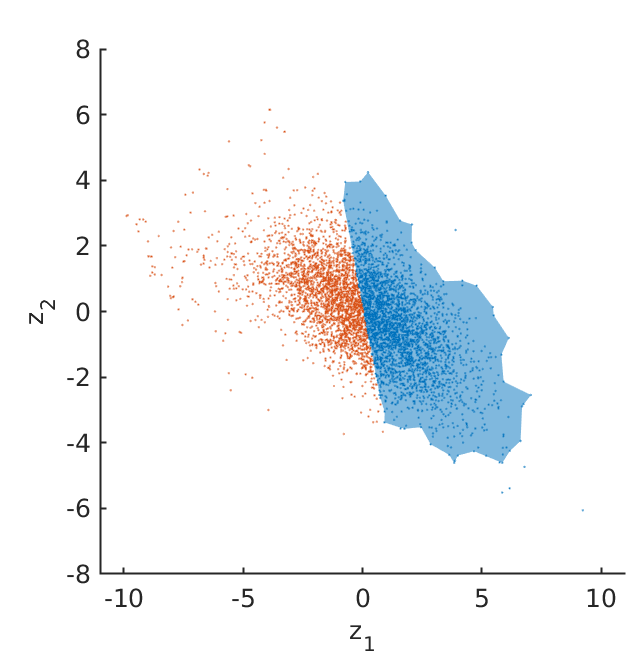}
    \caption{The area of the buggy region.}
    \label{fig:areabuggy}
\end{figure}

Given the instance space, the last step is to calculate the TISA metrics. The key metrics proposed in~\cite{neelofar24ICSE} are:

\begin{itemize}
    \item Area of the instance space, which refers to the region encompassed by all test instances within the instance space, estimating the overall diversity of the entire test suite.
    \item Area of the buggy region, which pertains to the section of the instance space taken up by most of the test instances that expose failures. An examples is shown in Figure~\ref{fig:areabuggy}
    \item Coverage of the instance space, which refers to the proportion of the total area as indicated by the boundary, that is covered by test instances. 

\end{itemize}

The experimental evaluation of these metrics showed that all TISA metrics demonstrate a positive correlation with faults, but the metric that measures the area of the buggy region (illustrated in Figure~\ref{fig:areabuggy}) particularly stands out due to its consistently robust and statistically significant correlation with bugs. This emphasizes its efficacy and reliability in providing valuable insights into the testing process's effectiveness.

In the future, there are opportunities in using TISA metrics to evaluate how well GenAI systems have been tested. One of the challenges is how to select a meaningful set of features to characterise such systems. One potential solution is to extract features from the embeddings of these systems, which contain rich semantic information by modelling the meaningful relationships and associations between words, captured by the numerical vectors used to represent those words in a high-dimensional space. Words with similar meanings or that are used in similar contexts will have word embeddings that are closer together in the vector space. Some recent word embeddings, like those produced by BERT~\cite{devlin2018bert}, incorporate contextual information. This means that the meaning of a word in a particular sentence depends on the surrounding words. This enables embeddings to capture more nuanced semantic information. 

In addition, embeddings often exhibit semantic relationships through analogies. Consider the words ``Paris'', ``France'', ``Rome'' and ``Italy''. Lets perform the following analogy by subtracting the vector for ``France'' from  ``Paris''. Then we add the vector for ``Italy''. The resulting vector is likely to be closest to the word ``Rome''. This analogy reflects the semantic relationship that ``Paris'' is to ``France'' as ``Rome'' is to ``Italy'' in terms of capital cities and their respective countries. It demonstrates how embeddings can capture geographical and cultural associations between words, providing another example of their ability to represent meaningful semantic information. All this information can be used to characterise the test cases used for testing a GenAI system, and can help explain why a test case is effective or ineffective. 

By constructing the instance space of the test scenarios used to test a GenAI system, we would be able to extract insights in terms of the diversity of the test inputs, and ensure that the test instances cover a wide range of scenarios, styles, or outputs that the GenAI system is expected to handle. This diversity helps assess the system's adaptability to various inputs. 

TISA would also help identify test instances of varying complexity levels, from simple and straightforward cases to more intricate and challenging scenarios. This helps assess the system's robustness and capability to handle complex inputs. In addition, TISA can be used to assess the realism of test cases, ensuring that test instances resemble real-world data or scenarios the AI system will encounter. 

Edge cases, outliers, or inputs that might trigger unusual behaviour in the AI system are important. Evaluating how well the system handles these cases can provide insights into its limitations and vulnerabilities. Introducing novel or unseen inputs to evaluate the system's ability to generalise beyond its training data and produce creative outputs is of particular importance, and can be tackled with a framework like TISA. By drawing the boundary of possible test inputs, and identifying areas of instance space where test inputs are sparse and effective (outliers), close to the boundary (edge cases), or close to the frontier of behaviours (unusual behaviours), TISA can help address these research challenges. 

To apply TISA to GenAI systems, there are a set of research challenges and opportunities for further research. First, a method that extracts features from test cases used to test GenAI systems would need to be developed. Let's take an example from other AI-based systems such as self driving cars. Features of a test case may constitute the number of other cars in the road, the number of pedestrians crossing the road, the weather conditions, the speed limit, the curvature of the road, etc. For GenAI systems such as question answering systems, a test case is text. Potential features could be extracted from the vector embeddings of the text, which contains rich semantic information. Such features would help characterise the test cases, and help the creation of the instance space. Another research opportunity is to explore how TISA can help with prioritising test cases, by enabling the selection of the test cases which are more likely to reveal bugs, thus reducing the overall testing time.

\section{Conclusion}

As the field of software testing encounters the novel challenges posed by GenAI systems, the need for new testing approaches becomes evident.
The Oracle problem, which involves establishing the correctness of creative outputs, stands as a central challenge in testing GenAI systems. The absence of a definitive ground truth, coupled with the subjectivity of human evaluators, has made it challenging to estimate the quality and correctness of generated content. The proposed solution of training a AI model to detect bias and deviations from the expected behaviour is a promising direction to mitigate this challenge.

Furthermore, addressing the adequacy of testing through measures like Test suite Instance Space Adequacy (TISA) metrics offers a quantitative and quantitative approach to assessing the diversity and coverage of test instances. By providing a two-dimensional representation of the instance space, TISA enables a systematic evaluation of the test suite's quality, revealing gaps and areas that require further testing.

As GenAI systems continue to permeate various domains, from creative content generation to complex decision-making, ensuring their reliability, robustness, and correctness is of paramount importance. This calls for a reimagining of testing methodologies that account for the inherent uncertainties and complexities of GenAI outputs.

\bibliographystyle{plain}
\bibliography{references.bib}
\end{document}